\documentclass[aps,prl,twocolumn,groupedaddress,showpacs]{revtex4-1}

\usepackage{amsmath}
\usepackage{graphicx}
\usepackage{bm}
\usepackage{amssymb}
\usepackage{float}

\usepackage{color}

\begin{document}

\title{Phase Transitions of Ferromagnetic Potts Models on the Simple Cubic
Lattice}

\author{S. Wang,$^1$ Z. Y. Xie,$^1$ J. Chen,$^1$ B. Normand,$^2$ and
T. Xiang$^1$}

\affiliation{$^1$Institute of Physics, Chinese Academy of Sciences,
P.O. Box 603, Beijing 100190, China}

\affiliation{$^2$Department of Physics, Renmin University of China,
Beijing 100872, China}

\date{\today}

\begin{abstract}

We investigate the 2- and 3-state ferromagnetic Potts models on the simple
cubic lattice using the tensor renormalization group method with higher-order
singular value decomposition (HOTRG). HOTRG works in the thermodynamic limit,
where we use the $Z_q$ symmetry of the model, combined with a new measure for
detecting the transition, to improve the accuracy of the critical point for
the 2-state model by two orders of magnitude, obtaining $T_c =
4.51152469(1)$. The 3-state model is far more complex, and we improve
the overall understanding of this case by calculating its thermodynamic
quantities with high accuracy. Our results verify the first-order nature
of the phase transition and the HOTRG transition temperature benchmarks
the most recent Monte Carlo result.

\end{abstract}


\maketitle

The phase transitions of the Potts model have for decades served as an
important paradigm for the study of critical properties, not only in hard
and soft condensed matter physics but also in fields as diverse as high-energy
physics and biophysics \cite{Potts,wu}. While the majority of these studies
have been for systems in two spatial dimensions, the three-dimensional (3D)
Potts model also contains independent and valuable insight for many problems.
As examples, the three-state ($q = 3$) 3D ferromagnetic Potts model plays
an important role in describing the finite-temperature deconfining
phase transition and the structure of quantum chromodynamics
(QCD) \cite{qcd,Berg,Baza}, while the antiferromagnetic Potts model
with $q \geqslant 3$ can be used to study entropy-driven phase transitions
in 3D \cite{Banavar,entropy,wang,Rosengren}.

The nature of the phase transition that occurs in the 3D three-state
ferromagnetic Potts model was in the past the subject of extended
controversy \cite{Ditzian,Straley,Enting,Herrmann}.
Most authors now favor a weakly first-order
transition \cite{BloteSwensdsen,Jensen,Fukugita,Lee,Janke,Nishino,Baza},
although a rigorous argument remains absent. In fact it is currently believed
that the 3D $q$-state ferromagnetic Potts model possesses a first-order phase
transition for all $q \geqslant 3$ and a continuous transition only when
$q = 2$ (the Ising case) \cite{Nishino,Bazavov}. The latent heat, which
measures the strength of the first-order nature, grows with $q$ \cite{Bazavov}.

The tensor renormalization group (TRG) method \cite{trg,srg,zhao,xie} is a
type of coarse-graining real-space renormalization technique, and continues
to draw increasing interest in condensed matter, statistical, and computational
physics. One of its primary advantages is that it is intrinsically in the
thermodynamic limit, yielding direct and highly accurate results with no
need for finite-size scaling (as in density matrix renormalization group
(DMRG) \cite{white}, Monte Carlo, and other numerical techniques). TRG-based
methods have been studied systematically \cite{srg,zhao,xie} and applied
with considerable success to statistical spin models
\cite{Chen2011,Meurice,Kadanoff,Yu}, gauge models \cite{yliu,Denbleyker},
and even quantum lattice models \cite{Jiang,Gu2008,Zhao2012,Xie2014}.

In this paper, we employ the recently developed TRG method based on
higher-order singular value decomposition (HOSVD), abbreviated as
HOTRG \cite{xie}, to study the thermodynamic properties of the 2- and
3-state ferromagnetic Potts models on the simple cubic lattice. The
general ferromagnetic $q$-state Potts model \cite{Potts}, which may be
considered as an extension of the Ising model to more than two components,
is defined by the Hamiltonian
\begin{equation}
H = - \sum_{\langle ij \rangle} \delta_{s_{i}s_{j}},
\end{equation}
where the sum is over all nearest-neighbor lattice sites and $s_{i} = 0,\ 1,\
\dots,\ q - 1$ denotes the $q$ different Potts states on site $i$.

\begin{figure}[t]
\centering
\includegraphics[width=0.45\textwidth]{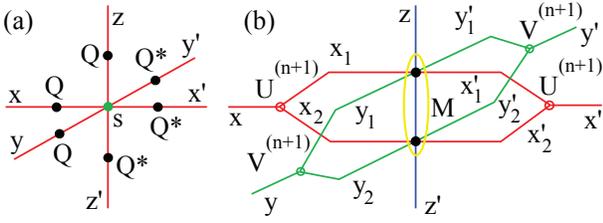}
\caption{(a) A local tensor is defined by summing over the common index $s$.
(b) Renormalization step updating the local tensor by $z$-axis contraction.}
\label{tensor}
\end{figure}

The partition function of any classical statistical model with only local
interactions, a category to which the Potts model belongs, can always be
represented by a tensor-network model \cite{trg,zhao}. To construct this
tensor network, we first expand the Boltzmann factor as
\begin{equation}
e^{\beta\delta_{s_{i}s_{j}}} = \sum_{t=0}^{q-1}Q_{s_{i} t}Q_{s_{j} t}^{*},
\nonumber
\end{equation}
where
\begin{equation}
Q_{s_{i} t}  =   e^{i 2\pi ts_{i}/q} \sqrt{\frac{e^{\beta}-1+q\delta_{t,0}}{q}},
\label{eq:q}
\end{equation}
and the superscript $(*)$ denotes complex conjugation. The local tensor may
then be defined \cite{zhao} as
\begin{equation}
T_{x_{i}x_{i}'y_{i}y_{i}'z_{i}z_{i}'} = \sum_{s_{i}} Q_{s_{i} x_{i}} Q^{*}_{s_{i} x_{i}'}
Q_{s_{i} y_{i}} Q^{*}_{s_{i} y_{i}'} Q_{s_{i} z_{i}} Q^{*}_{s_{i} z_{i}'}, \label{eq:T}
\end{equation}
where $(x_{i},\ x_{i}',\ y_{i},\ y_{i}',\ z_{i},\ z_{i}' )$
are respectively the tensor indices for the $x$, $y$, and $z$ directions,
and is represented schematically in Fig.~\ref{tensor}(a). The partition
function is represented in terms of these tensors by
\begin{equation}
Z = \text{Tr} \prod_{i} T_{x_{i}x_{i}'y_{i}y_{i}'z_{i}z_{i}'},
\end{equation}
where the trace is over all repeated indices.

In 2D, tensor-network models can be evaluated easily by using TRG methods
that contract the network efficiently. However, the evaluation of 3D tensor
networks was an intractable problem, in terms of computational effort and
accuracy, until the proposal of HOTRG \cite{xie}. Specializing our description
to the simple cubic lattice, we assume for simplicity that all local tensors
$T$ are identical. At the $n$th renormalization step of HOTRG, one first
contracts two adjacent tensors with a common index $i$ along the
$z$-direction to form a new tensor $M$,
\begin{equation}
M_{ijklzz'}^{(n)} = \sum_{\widetilde{z}} T_{x_{1}x'_{1}y_{1}y'_{1}z \widetilde{z}}^{(n)}
T_{x_{2}x_{2}'y_{2}y_{2}'\widetilde{z} z'}^{(n)},
\end{equation}
where $i = x_{1} \otimes x_{2}$, $j = x_{1}' \otimes x_{2}'$, $k = y_{1} \otimes
y_{2}$, and $l = y_{1}'\otimes y_{2}'$. Clearly the bond dimension of the new
tensor in the $x$ and $y$ directions is the square of that in the original
tensor, and a truncation scheme is necessary to avoid an exponential
divergence.

To perform an optimal truncation, HOTRG employs the HOSVD to decompose $M^{(n)}$
as
\begin{equation}
M_{ijklzz'}^{(n)} \! = \!\! \sum_{xx'yy'mn} \!\!\! S_{xx'yy'mn} U_{ix}^{L} U_{jx'}^{R}
U_{ky}^{F} U_{ly'}^{B} U_{zm}^{U} U_{z'n}^{D},
\end{equation}
where the matrices $U$ are all unitary and $S$ is known as the core
tensor, one with the properties of full orthogonality and
pseudo-diagonality \cite{hosvd}. Details of the practical determination
of the six unitary matrices may be found in Ref.~[\onlinecite{xie}].
Next one compares the values of
\begin{equation}
\varepsilon_{1} = \sum_{i>D}|S_{i,:,:,:,:,:}| \;\; {\rm and} \;\;
\varepsilon_{2} = \sum_{i>D}|S_{:,i,:,:,:,:}| ,
\end{equation}
where $D$ is the bond dimension retained after truncation.
\[
|S_{i,:,:,:,:,:}| = \sqrt{ \sum_{x'yy'zz'} S_{ix'yy'zz'}^2 }
\]
is the norm of the subtensor $S_{i,:,:,:,:,:}$  and $|S_{:,i,:,:,:,:}|$ is defined
similarly. If $\varepsilon_{1} < \varepsilon_{2}$ (or $\varepsilon_{1} >
\varepsilon_{2}$), one truncates the second dimension of $U^{L}$ (or $U^{R}$)
to $D$ to form an isometry $U^{(n+1)}$. A similar treatment applied to the
$y$-direction yields another isometry $V^{(n+1)}$, with which the renormalized
local tensor is then updated as
\begin{equation}
T_{xx'yy'zz'}^{(n+1)} \! = \! \sum_{ijkl} \! M_{ijklzz'}^{(n)} U_{ix}^{(n+1)}
U_{jx'}^{(n+1)} V_{ky}^{(n+1)} V_{ly'}^{(n+1)}.
\end{equation}
The schematic representation of this process is
shown in Fig.~\ref{tensor}(b). This type of truncation scheme provides a good
local approximation to minimize the truncation error and to conserve the
optimal amount of information about the system. In practical calculations,
the lattice is contracted along the $x$, $y$, and $z$ directions in sequence
until the desired quantities have converged.

Clearly the size of the lattice is reduced by a factor of $2$ after an HOTRG
step. Alternatively stated, $n$ HOTRG steps represent a system of size $2^{n}$,
and with sufficiently large $n$ it is easy to approach the thermodynamic
limit. In practice $n = 30$ is enough for most systems. This sort of size is
inordinately difficult to reach by other methods such as Monte Carlo, and it
is this convenience with which HOTRG can access the infinite system, instead
of requiring a finite-size scaling analysis, which is one of its prime
advantages over other methods. In general, the accuracy of HOTRG is limited
by the local truncation error and the bond dimension $D$ retained during the
renormalization step. In 3D, the computational complexity and the memory cost
of the HOTRG algorithm scale respectively as $D^{11}$ and $D^{6}$ \cite{xie}.

One means of increasing the accessible bond dimension is to make full use of
the symmetries of the Hamiltonian, an approach employed in all numerical methods
to eliminate unnecessary memory use and redundant computation \cite{zhao}, thus
reducing the computational cost quite significantly. The $q$-state Potts model
possesses $Z_{q}$ symmetry and it is straightforward to show that the elements
of the initial local tensor $T_{x_{i}x_{i}'y_{i}y_{i}'z_{i}z_{i}'}$ (\ref{eq:T}) are
nonzero only when the indices satisfy the relation
\begin{equation}
\mathrm{mod}( x_{i} + y_{i} + z_{i}, q ) = \mathrm{mod} (x_{i}' + y_{i}' + z_{i}',
q).
\end{equation}
One may further verify that this relation is maintained throughout the
renormalization procedure, which is of vital importance in practical
calculations.

\begin{figure}[t]
\centering
\includegraphics[width=0.38\textwidth]{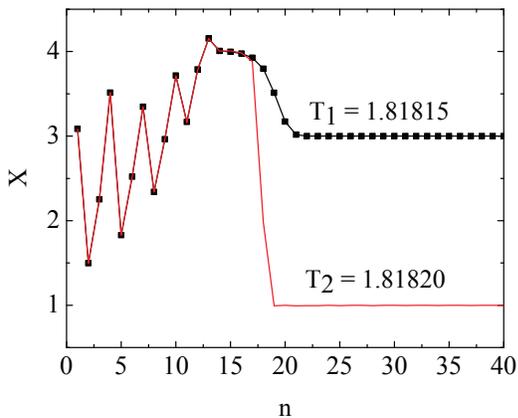}
\caption{Evolution of $X$ with the number $n$ of RG steps in HOTRG
calculations for the $q = 3$ Potts model with $D = 16$. Values are shown
for two temperatures $T_{1} = 1.81815$ and $T_{2} = 1.81820$, which are in
the ordered and disordered phases respectively. Although $T_{1}$ and $T_{2}$
are very close, $X$ approaches one of two explicitly different values, $q$
or 1, when $n > 20$, indicating clearly that a phase transition occurs at
$T_1 < T < T_2$.}
\label{flow}
\end{figure}

We seek an accurate measure appropriate to the TRG scheme to fix the
transition point explicitly. After sufficiently many renormalization steps,
$T$ approaches a fixed-point tensor which behaves differently in different
phases. A quantitative measure for this difference, introduced in
Ref.~[\onlinecite{Gu2009}], is
\begin{equation}
X = \frac{(\text{Tr} A)^{2}}{\text{Tr}(A^2)} ,
\label{eq:X}
\end{equation}
where $A$ is a $D\times D$ matrix defined by
\begin{equation}
A_{z_{i}z_{i}'} = \sum_{x_{i}y_{i}} T_{x_{i}x_{i}y_{i}y_{i}z_{i}z_{i}'}.
\label{eq:e5}
\end{equation}
For the fixed point tensor $T$ in the high-temperature disordered phase,
all eigenvalues of $A$ are close to zero other than the first, whereas in
the low-temperature symmetry-breaking phase the first $q$ eigenvalues are
approximately equal with the others zero. Thus the degeneracy of the largest
eigenvalue of $A$ can be used as an indicator of symmetry-breaking, and is
given by the value of $X$. Figure \ref{flow} shows the behavior of $X$ for
the $q = 3$ Potts model during the renormalization group (RG) flow at two
temperatures differing by only $5 \times 10^{-5}$. The almost exact
convergence of $X$ to the values $q$ or 1 beyond $n = 20$ illustrates
the power of this approach.

\begin{table}[b]
\caption{Comparison of critical temperatures $T_{c}$ obtained by different
methods for the Ising model on the simple cubic lattice.}
\centering{}%
\begin{tabular}{llllllllll}\\
\hline
\hline
Method &  &  &  &  &  &  &  &  & \multicolumn{1}{c}{$T_{c}$}\tabularnewline
\hline
CTMRG (2001) \cite{Nishin} &  &  &  &  &  &  &  &  & 4.5393
\tabularnewline
TPVA (2005) \cite{Gendiar} &  &  &  &  &  &  &  &  & 4.554
\tabularnewline
Algebraic variation (2006) \cite{Chung} &  &  &  &  &  &  &  &  & 4.547
\tabularnewline
Series expansion (2000) \cite{Butera} &  &  &  &  &  &  &  &  & 4.511536(21)
\tabularnewline
Monte Carlo RG (1996) \cite{Gupta} &  &  &  &  &  &  &  &  & 4.5115(2)
\tabularnewline
Monte Carlo (2003) \cite{Deng} &  &  &  &  &  &  &  &  & 4.5115248(6)
\tabularnewline
Monte Carlo (2010) \cite{Hasenb} &  &  &  &  &  &  &  &  & 4.5115232(17)
\tabularnewline
HOTRG ($D = 16$) (2012) \cite{xie} &  &  &  &  &  &  &  &  & 4.511544
\tabularnewline
HOTRG ($D = 23$, this work) &  &  &  &  &  &  &  &  & 4.51152469(1)
\tabularnewline
\hline
\hline
\end{tabular}\label{T-Ising}
\end{table}

\begin{figure}[t]
\centering
\includegraphics[width=0.38\textwidth]{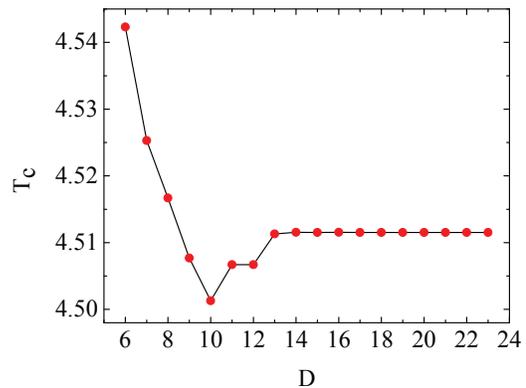}
\caption{Critical temperature $T_{c}$ as a function of the bond dimension $D$
obtained by HOTRG for the Ising model on the simple cubic lattice.}
\label{F-Ising}
\end{figure}

The HOTRG method was first applied to study the 3D Ising model (the $q = 2$
Potts model on the simple cubic lattice) in Ref.~[\onlinecite{xie}]. There
the maximum attainable bond dimension was $D = 16$, with which the accuracy
of the critical temperature $T_c$ was comparable to the best Monte Carlo
result. In this work, by employing the $Z_q$ symmetry, we increase $D$ to
$23$. As shown in Fig.~\ref{F-Ising}, the critical temperature $T_c$ converges
very accurately beyond $D = 13$, to the value $T_c = 4.51152469(1)$. Table
\ref{T-Ising} shows the comparison with results obtained for $T_c$ in this
model by other methods. To our knowledge, our result is two orders of magnitude
more precise than the best alternative treatment to date (Monte Carlo) and six
orders more precise than the results obtained by the corner transfer matrix
renormalization group (CTMRG) and the tensor product variational approach
(TPVA) introduced by Nishino and coworkers \cite{Nishino,Nishin,Gendiar}. This
result displays both the power of the $Z_q$ symmetry and the accuracy of the
HOTRG algorithm at higher $D$, even compared to other tensor-based approaches.

\begin{figure}[t]
\centering
\includegraphics[width=0.38\textwidth]{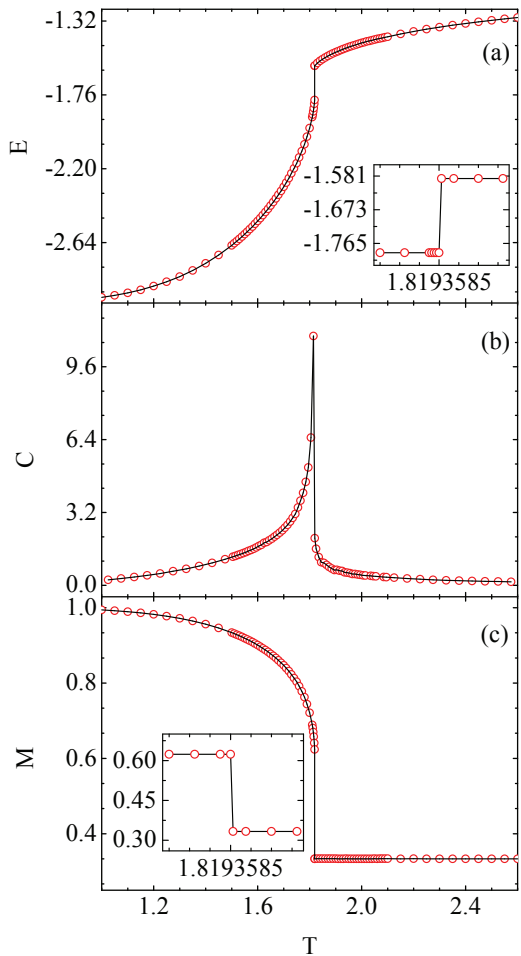}
\caption{(a) Internal energy per site, (b) specific heat, and (c) magnetization
computed as functions of temperature by HOTRG for the $q = 3$ ferromagnetic
Potts model on the simple cubic lattice with bond dimension $D = 14$. Insets
in panels (a) and (c) show respectively the discontinuities in internal energy
and magnetization at the phase transition which indicate its first-order
nature.}
\label{Potts}
\end{figure}

We now turn to the 3-state Potts model, which is significantly more challenging
and less well understood. We first illustrate the thermodynamic quantities
computed by HOTRG, for which $D = 14$ provides representative results (as
shown below). We have calculated the internal energy $E$, the specific heat
$C$, and the ferromagnetic magnetization $M = \sum_{i} \delta_{s_{i},0} / N$.
In Fig.~\ref{Potts}(b) one observes a sharp peak in the specific-heat curve,
indicative of a phase transition. That this transition has first-order nature
is demonstrated by the corresponding discontinuity in the energy curve around
$T = 1.819$, shown in Fig.~\ref{Potts}(a). With a temperature resolution of
$10^{-5}$ [inset, Fig.~\ref{Potts}(a)], the latent heat $\Delta E$, which
denotes the energy difference between the two phases at the critical point,
is clearly finite, with a value $\Delta E = 0.2029$ for $D = 14$. A comparison
of the results obtained for this quantity by different methods is shown in
Table \ref{Latent}; while the HOTRG result is fully consistent with the spread
of available values, we make no claims concerning the quantitative accuracy
of the $D = 14$ value, and compute it primarily to confirm the qualitative
nature of the transition.

\begin{figure}
\centering
\includegraphics[width=0.38\textwidth]{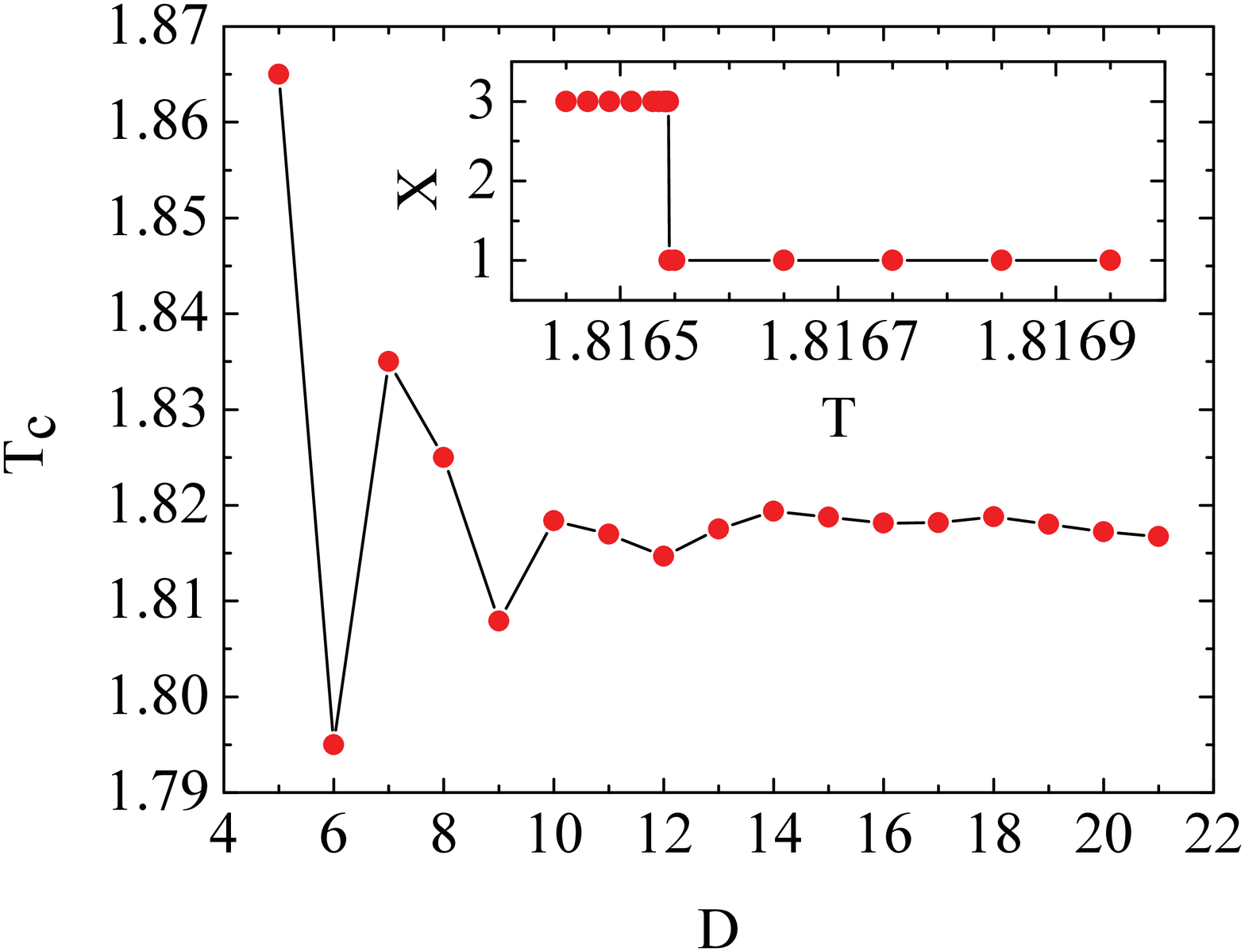}
\caption{Transition temperature $T_{c}$ as a function of bond dimension $D$
for the $q = 3$ ferromagnetic Potts model on the simple cubic lattice. Inset:
behavior of $X$ (Eq.\ref{eq:X}) as a function of temperature for $D = 21$, where
the abrupt step in the curve gives a phase transition at $T_{c} = 1.8165945$.}
\label{Tcd}
\end{figure}

\begin{table}[b]
\caption{Comparison of latent heats $\Delta E$ and transition temperatures
$T_{c}$ for the $q = 3$ ferromagnetic Potts model on the simple cubic lattice
obtained by different methods. The tensor bond dimensions retained in the
HOTRG calculations of $\Delta E$ and $T_{c}$ were respectively $D = 14$ and
$D = 21$. $L$ denotes the largest system size (cube side) reached in the
Monte Carlo simulations.}
\centering{}%
\begin{tabular}{lll}\\
\hline
\hline
Method & $\Delta E$ &$T_{c}$
\\
\hline
Series expansion (1979) \cite{Miyashita} &  & 1.7289(12) \\
Monte Carlo RG (1979) \cite{BloteSwensdsen} &  & 1.818 \\
Monte Carlo (1982, $L = 8$) \cite{Ito} & 0.12 & 1.81 \\
Pair approximation (1982) \cite{Ito} & 0.123 & 1.879 \\
Monte Carlo (1987, $L = 16$) \cite{Wilson} & 0.2222(7) & 1.81618(7)
\\
Monte Carlo (1991, $L = 36$) \cite{Alves} \quad & 0.16062(52) & 1.816455(35)
\\
Monte Carlo (1997, $L = 36$) \cite{Janke} & 0.1614(3) & 1.816316(33)
\\
Monte Carlo (2007, $L = 50$) \cite{Baza} & 0.1643(8) & 1.816315(19)
\\
TPVA (2002) \cite{Nishino} & 0.228 & 1.8195 \\
HOTRG (this work) & 0.2029  & 1.8166  \\
& ($D = 14$) & ($D = 21$) \\
\hline
\hline
\end{tabular}\label{Latent}
\end{table}

Figure \ref{Potts}(c) shows the behavior of the magnetization, which decreases
smoothly from 1 until $T$ approaches the critical point, where it falls sharply
to a value of $1/3$. When $T = 0$, the ferromagnetic interaction aligns all the
spins and $M = 1$. As $T$ grows, increasing thermal fluctuations cause the
steady decrease, but the system remains in the symmetry-broken phase. Only
when $T > T_c$ does the system enter the fully disordered phase, where
$M = 1/3$ (all spin components equal). Once again the first-order nature
of the transition is clearly visible in the sharp drop of the order parameter
at $T_c$ [inset, Fig.~\ref{Potts}(c)], which is $\Delta M = 0.2903$ for $D =
14$.

A more accurate estimate of $T_c$ is essential for the study of critical
properties, and for this we exploit the symmetry of the tensor-network model
to raise the bond dimension to $D = 21$. From the temperature dependence of
$X$ we obtain $T_c = 1.8165945$ for $D = 21$ [inset, Fig.~\ref{Tcd}].
However, because the accuracy of HOTRG improves at larger bond dimensions
(Fig.~\ref{F-Ising}), it is necessary to investigate the behavior of $T_c$
with $D$. We show in Fig.~\ref{Tcd} the convergence of $T_c$ for the $q = 3$
case, where it is clear that the results are not yet in the convergent
regime for any accessible $D$ values. However, because they have almost
converged for $D = 21$, we are currently able to estimate the true transition
temperature and error bar as $T_{c} = 1.8166(5)$, which sets a valuable
additional benchmark. A comparison of $T_c$ obtained by different methods
is shown in Table \ref{Latent}, where the $D = 21$ HOTRG result agrees
well (relative error $10^{-4}$) with the most recent Monte Carlo
simulations \cite{Baza}.

On the other hand, the lack of convergence even by the most sophisticated
HOTRG methods and measures indicates that the 3-state Potts model is a
genuinely hard problem. To achieve the same type of convergence as in the
Ising model, within the same calculational framework, requires a still
larger tensor dimension $D$, which is computationally intensive in both
time and memory. An alternative approach is to include the effect of
the bond environment for a global optimization of the truncation during the
coarse-graining process, as in the highly efficient second renormalization
group method \cite{srg,xie}. By comparison with previous work for both the
$q = 2$ and 3 Potts models, it is clear that the HOTRG method already
offers comparable results for the challenging $q = 3$ case (Table II)
and that it has very significant potential for improvement, as already
achieved for the $q = 2$ case (Table I).

In summary, we have investigated the 2- and 3-state ferromagnetic Potts
models on the simple cubic lattice by using the recently developed tensor
renormalization group technique, which can study the thermodynamic limit
directly with no need for finite-size scaling. By employing the refined
HOTRG method, exploiting the $Z_{q}$ symmetry of the Hamiltonians, and
introducing a TRG-specific measure for the ground-state properties, we
have determined the phase-transition temperature to high accuracy.
For the 2-state model with $D = 23$, we have obtained, to our knowledge, by
far the most accurate critical temperature of any available method, $T_c =
4.51152469(1)$. For the 3-state model with $D = 14$, we have calculated the
thermodynamic quantities with high precision, specifically the internal
energy, latent heat, specific heat, and magnetization. These results
verify the first order nature of the symmetry-breaking phase transition.
By reaching $D = 21$ we obtain a phase-transition temperature $T_{c} =
1.8166$, which is close to the extrapolated Monte Carlo result.

\acknowledgments

This work was supported by the National Natural Science Foundation of China
(Grant Nos.~10934008, 10874215, and 11174365) and by the National Basic
Research Program of China (Grant Nos.~2012CB921704 and 2011CB309703).


\begin{thebibliography}{10}


\bibitem{Potts} R. B. Potts, Proc. Cambridge. Philos. Soc. \textbf{48},
106 (1952).

\bibitem{wu} F. Y. Wu, Rev. Mod. Phys. \textbf{54}, 235 (1982).

\bibitem{qcd} B. Svetitsky and L. G. Yaffe, Nucl. Phys. B \textbf{210}, 423
(1982).

\bibitem{Berg} B. A. Berg, H. Meyer-Ortmanns, and A. Velytsky, Phys. Rev. D
\textbf{70}, 054505 (2004).

\bibitem{Baza} A. Bazavov and B. A. Berg, Phys. Rev. D \textbf{75}, 094506
(2007).

\bibitem{Banavar} J. R. Banavar, G. S. Grest, and D. Jasnow, Phys. Rev. Lett.
\textbf{45}, 1424 (1980); Phys. Rev. B \textbf{25}, 4639 (1982).

\bibitem{entropy} Y. Ueno, G. Sun, and I. Ono, J. Phys. Soc. Jpn. \textbf{58},
1162 (1989).

\bibitem{wang} S. Wang, R. H. Swendsen, and R. Kotecky, Phys. Rev. Lett.
\textbf{63}, 109 (1989); Phys. Rev. B \textbf{42}, 2465 (1990).

\bibitem{Rosengren} A. Rosengren and S. Lapinskas, Phys. Rev. Lett.
\textbf{71}, 165 (1993).

\bibitem{Ditzian} R. V. Ditzian and J. Oitmaa, J. Phys. A \textbf{7}, 16 (1974).

\bibitem{Straley} J. P. Straley, J. Phys. A \textbf{7}, 2173 (1974).

\bibitem{Enting} I. G. Enting, J. Phys. A \textbf{7}, 1617 (1974).

\bibitem{Herrmann} H. J. Herrmann, Z. Phys. B \textbf{35}, 171 (1979).

\bibitem{BloteSwensdsen} H. W. J. Blote and R. H. Swendsen, Phys. Rev. Lett.
\textbf{43}, 799 (1979).

\bibitem{Jensen} S. J. K. Jensen and O. G. Mouritsen, Phys. Rev. Lett.
\textbf{43}, 1736 (1979).

\bibitem{Fukugita} M. Fukugita and M. Okawa, Phys. Rev. Lett. \textbf{63},
13 (1989).

\bibitem{Lee} J. Lee and J. M. Kosterlitz, Phys. Rev. B \textbf{43}, 1268
(1991).

\bibitem{Janke} W. Janke and R. Villanova, Nucl. Phys. B \textbf{489}, 679
(1997).

\bibitem{Nishino} T. Nishino, K. Okunishi, Y. Hieida, N. Maeshima, and Y.
Akutsu, Nucl. Phys. B \textbf{575}, 504 (2000); A. Gendiar and T. Nishino,
Phys. Rev. E \textbf{65}, 046702 (2002).

\bibitem{Bazavov} A. Bazavov, B. A. Berg, and S. Dubey, Nucl. Phys. B
\textbf{802}, 421 (2008).

\bibitem{trg} M. Levin and C. P. Nave, Phys. Rev. Lett. \textbf{99}, 120601
(2007).

\bibitem{srg} Z. Y. Xie, H. C. Jiang, Q. N. Chen, Z. Y. Weng, and T. Xiang,
Phys. Rev. Lett. \textbf{103}, 160601 (2009).

\bibitem{zhao} H. H. Zhao, Z. Y. Xie, Q. N. Chen, Z. C. Wei, J. W. Cai,
and T. Xiang, Phys. Rev. B \textbf{81}, 174411 (2010).

\bibitem{xie} Z. Y. Xie, J. Chen, M. P. Qin, J. W. Zhu, L.P. Yang, and
T. Xiang, Phys. Rev. B \textbf{86}, 045139 (2012).

\bibitem{white} S. R. White, Phys. Rev. Lett. \textbf{69}, 2863 (1992).

\bibitem{Chen2011} Q. N. Chen, M. P. Qin, J. Chen, Z. C. Wei, H. H. Zhao,
B. Normand, and T. Xiang, Phys. Rev. Lett. \textbf{107}, 165701 (2011).

\bibitem{Meurice} Y. Meurice, Phys. Rev. B \textbf{87}, 064422 (2013).

\bibitem{Kadanoff} E. Efrati, Z. Wang, A. Kolan, and L. P. Kadanoff,
unpublished (arXiv:1301.6323).

\bibitem{Yu} J. F. Yu, Z. Y. Xie, Y. Meurice, Y. Liu, A. Denbleyker, H. Zou,
M. P. Qin, J. Chen, and T. Xiang, Phys. Rev. E \textbf{89}, 013308 (2014).

\bibitem{yliu} Y. Liu, Y. Meurice, M. P. Qin, J. U. Yockey, T. Xiang,
Z. Y. Xie, J. F. Yu, and H. Zou, Phys. Rev. D \textbf{88}, 056005 (2013).

\bibitem{Denbleyker} A. Denbleyker, Y. Liu, Y. Meurice, M. P. Qin, T. Xiang,
Z. Y. Xie, J. F. Yu, and H. Zou, Phys. Rev. D \textbf{89}, 016008 (2014).

\bibitem{Jiang} H. C. Jiang, Z. Y. Weng, and T. Xiang, Phys. Rev. Lett.
\textbf{101}, 090603 (2008).

\bibitem{Gu2008} Z. C. Gu, M. Levin, and X. G. Wen, Phys. Rev. B \textbf{78},
205116 (2008).

\bibitem{Zhao2012} H. H. Zhao, C. Xu, Q. N. Chen, Z. C. Wei, M. P. Qin,
G. M. Zhang, and T. Xiang, Phys. Rev. B \textbf{85}, 134416 (2012).

\bibitem{Xie2014} Z. Y. Xie, J. Chen, J. F. Yu, X. Kong, B. Normand, and
T. Xiang, Phys. Rev. X \textbf{4}, 011025 (2014).

\bibitem{hosvd} L. de Lathauwer, B. de Moor, and J. Vandewalle, SIAM J. Matrix
Anal. Appl. \textbf{21}, 1253 (2000).

\bibitem{Gu2009} Z. C. Gu and X. G. Wen, Phys. Rev. B \textbf{80}, 155131
(2009).

\bibitem{Nishin} T. Nishino, Y. Hieida, K. Okunishi, N. Maeshima,
Y. Akutsu, and A. Gendiar, Prog. Theo. Phys, \textbf{105}, 409 (2001).

\bibitem{Gendiar} A. Gendiar and T. Nishino, Phys. Rev. B \textbf{71},
024404 (2005).

\bibitem{Chung} S. G. Chung, Phys. Lett. A \textbf{359}, 707 (2006).

\bibitem{Butera} P. Butera and M. Comi, Phys. Rev. B \textbf{62}, 14837 (2000).

\bibitem{Gupta} R. Gupta and P. Tamayo, Int. J. Mod. Phys. C \textbf{7},
305 (1996).

\bibitem{Deng} Y. J. Deng and H. W. J. Blote, Phys. Rev. E \textbf{68},
036125 (2003).

\bibitem{Hasenb} M. Hasenbusch, Phys. Rev. B \textbf{82}, 174433 (2010).

\bibitem{Ito} I. Ono and K. Ito, J. Phys. C \textbf{15}, 4417 (1982).

\bibitem{Wilson} W. G. Wilson and C. A. Vanse, Phys. Rev. B \textbf{36},
587 (1987).

\bibitem{Alves} N. A. Alves, B. A. Berg, and R. Villanova, Phys. Rev. B
\textbf{43}, 5846 (1991).

\bibitem{Miyashita} S. Miyashita, D. D. Betts, and C. J. Elliott, J. Phys. A
\textbf{12}, 1605 (1979).

\end{thebibliography}
\end{document}